\documentclass[twocolumn,showpacs,amsmath,amssymb,aps,prb,superscriptaddress,tbtags]{revtex4-1}

\usepackage[]{graphicx}
\usepackage[]{verbatim}
\usepackage[]{color}
\usepackage{overpic}
\usepackage{rotating}
\usepackage{mathtools}
\usepackage{appendix}
\usepackage{ulem}
\usepackage[margin=0.6in]{geometry}
\usepackage{gensymb}
\usepackage{tabularx}

\usepackage[]{hyperref}

 \hypersetup{
 colorlinks = true,
 linkcolor = blue,
 urlcolor = blue,
 citecolor = blue}

\usepackage{array}
\newcolumntype{L}[1]{>{\raggedright\let\newline\\\arraybackslash\hspace{0pt}}m{#1}}
\newcolumntype{C}[1]{>{\centering\let\newline\\\arraybackslash\hspace{0pt}}m{#1}}
\newcolumntype{R}[1]{>{\raggedleft\let\newline\\\arraybackslash\hspace{0pt}}m{#1}}
\newcolumntype{N}{@{}m{0pt}@{}}

\makeatletter
\newsavebox{\@brx}

\newcommand{\llangle}[1][]{\savebox{\@brx}{\(\m@th{#1\langle}\)}%
  \mathopen{\copy\@brx\mkern2mu\kern-0.8\wd\@brx\usebox{\@brx}}}
\newcommand{\rrangle}[1][]{\savebox{\@brx}{\(\m@th{#1\rangle}\)}%
  \mathclose{\copy\@brx\mkern2mu\kern-0.8\wd\@brx\usebox{\@brx}}}

  \newcommand{\lllangle}[1][]{\savebox{\@brx}{\(\m@th{#1\langle}\)}%
  \mathopen{\copy\@brx\copy\@brx\mkern4mu\kern-0.7\wd\@brx\usebox{\@brx}}}
\newcommand{\rrrangle}[1][]{\savebox{\@brx}{\(\m@th{#1\rangle}\)}%
  \mathclose{\copy\@brx\copy\@brx\mkern4mu\kern-0.7\wd\@brx\usebox{\@brx}}}

\graphicspath{{./}{./}}

\begin{document}
\title{Odd-Parity Superconductivity Driven by Octahedra Rotations in Iridium Oxides}
\author{Austin W.~Lindquist}
\affiliation{Department of Physics and Center for Quantum Materials, University of Toronto, 60 St.~George St., Toronto, Ontario, M5S 1A7, Canada}
\author{Hae-Young Kee}
\email{hykee@physics.utoronto.ca}
\affiliation{Department of Physics and Center for Quantum Materials, University of Toronto, 60 St.~George St., Toronto, Ontario, M5S 1A7, Canada}
\affiliation{Canadian Institute for Advanced Research, Toronto, Ontario, M5G 1Z8, Canada}
\begin{abstract}
Iridium oxides have provided a playground to study novel phases originating from spin-orbit coupling and electron-electron interactions.  
Among them, the d-wave singlet superconductor was proposed for electron-doped Sr$_2$IrO$_4$, containing two Ir atoms in a unit cell due to the staggered rotation of oxygen octahedra about the c-axis.  
It was also noted that such oxygen octahedra rotation affects electronic transports.
Here we study the role of 
octahedra tilting away from the c-axis, in determining superconducting pairing symmetry.
We show that the octahedra tilting 
changes the large Fermi surface to a Dirac point, which strongly suppresses the conventional d-wave pairing.
Furthermore, it also promotes effective spin-triplet interactions in the strong Hubbard interaction limit, 
leading to a transition from the even-parity to odd-parity superconducting phase.
Thus, tuning octahedra distortions can be used as a tool to engineer a spin triplet superconductor in strongly correlated systems with strong spin-orbit coupling.
\end{abstract}
\maketitle

\section{Introduction} 

Spin-orbit coupling (SOC) has provided a new avenue by which novel phases in strongly correlated electronic systems can be achieved \cite{Witczak2014,Rau2016}.  
This influences both the electronic and magnetic structure of the system, and has led to the proposal of phases such as a Kitaev spin liquid in the honeycomb iridium oxides (iridates), A$_2$IrO$_3$, and RuCl$_3$ \cite{Kitaev2006ap,Jackeli2009prl,Rau2014prl,Plumb2014prb,Kim2015prb,Banerjee2016nm,Schaffer2016rpp,Winter2017jpcm}, as well as unconventional Mott insulators \cite{Moon2008prl,Kim2008prl,Kim2009s} in a series of perovskite iridates, Sr$_{n+1}$Ir$_n$O$_{3n+1}$, 
and topological nodal line semimetals in SrIrO$_3$ \cite{Carter2012prb,Chen2015nc,Fang2015prb}.

These recent intense activities in strongly correlated systems with strong SOC were initially undertaken for the purpose of understanding the insulating behaviour of Sr$_2$IrO$_4$ \cite{Moon2008prl,Kim2008prl,Kim2009s}.
After taking into account the strong SOC in iridates, a Mott insulator with 
 the amplified Hubbard interaction on a narrow $J_{\text{eff}}=\frac{1}{2}$ bandwidth was proposed \cite{Moon2008prl,Kim2008prl,Kim2009s,Jin2009prb}.
Given that the relevant band is made of a single $J_{\text{eff}}=\frac{1}{2}$ orbital and a relatively strong Hubbard interaction \cite{Kim2008prl}, 
a proposed microscopic Hamiltonian is a half-filled, single band, Hubbard model, 
similar to that of the cuprates \cite{Lee2006rmp,Fujiyama2012prl}. It is interesting to note earlier efforts which tried to find a non-copper high-T$_c$ superconductor
on iridates, given the similarity of the crystal structure to such as La$_2$CuO$_4$ \cite{Crawford1994prb,Cao1998prb,Kim2012prl,Terashima2017prb}.  
%
Based on this similarity to the superconducting cuprates, $d$-wave superconductivity has been suggested in electron doped Sr$_2$IrO$_4$ \cite{Wang2011prl,Watanabe2013prl,Yang2014prb,Meng2014prl}.

While there is currently no experimental evidence of superconductivity \cite{Bertinshaw2019}, 
Fermi arcs have been observed in electron doped samples via angle-resolved photoemission spectroscopy (ARPES) measurements \cite{Kim2014s,DelaTorre2015prl}, hinting at a possible anisotropic pseudogap feature, shared in many other correlated materials \cite{Norman1998prb,Timusk1999rpp,Damascelli2003rmp,Yoshida2012}.  More recently, these Fermi arcs have been shown to shrink to point nodes as a $d$-wave symmetric gap opens at low temperatures \cite{Kim2016np}.  
Around the same time, the effect of distortions was studied experimentally in the layered perovskite iridate, Sr$_2$IrO$_4$ \cite{Korneta2010prb,Ge2011prb,Clancy2014prb,Cao2018rpp}.  
It was noted that transport behaviour significantly changes by introducing oxygen vacancies \cite{Korneta2010prb}, as well as by introducing electron (or hole) doping using La (or K) atoms \cite{Klein2008,Ge2011prb},
 all of which correlate to the rotation angles of the oxygen octahedra.  



Considering the strong impact of octahedra rotations on the electronic properties, here, we investigate if the superconducting pairing symmetry depends on these rotations.
There are two rotations involved in an octahedra cage. One is a staggered rotation denoted by the angle $\phi$ around the $c$-axis which occurs in Sr$_2$IrO$_4$ making two Ir atoms in a unit cell \cite{Crawford1994prb,Ye2013prb}, as shown in the lower inset of Fig.~\ref{BS}(a).
The other rotation is a tilting away from the $c$-axis denoted by the angle $\theta$, as shown in the lower inset of Fig.~\ref{BS}(c). 
To differentiate these two rotations, we will refer to the angle $\phi$ as the rotation, and the angle $\theta$ as the tilting from now on.

Below we show that the introduction of the tilting and rotation of the oxygen octahedra modifies the shape of the band structure to a Dirac Fermi surface protected by
a glide symmetry. This limits the Fermi surface to small pockets in the area where the even-parity paring gap is minimized, while the odd-parity pairing gap is maximized.
Furthermore, effective spin interactions in the strong Hubbard $U$ limit favouring both even- and odd-parity pairing are generated by the tilting, while the conventional Heisenberg interaction
favors only spin-singlet, even-parity pairing.
Thus the odd-parity superconducting phase can be found via increasing the octahedra tilting, originating from
a combination of Fermi surface change and additional spin interactions.
Both the tilting and rotation occur in the bulk SrIrO$_3$ \cite{Longo1971,Zhao2008jap}, and likely happen in an iridium oxide layer grown on a substrate, AMO$_3$, which has been done using pulsed laser deposition \cite{Matsuno2015prl},
where 
AMO$_3$ is a band insulator with a closed shell transition metal M
and a crystal structure including both tilting and rotation such as P$_{\text{bnm}}$.

This paper is organized as follows.  In Sec.~2, we show how the electronic band dispersion of $J_{\rm eff}=\frac{1}{2}$ changes
when a tilting angle, $\theta$, is introduced, while the glide symmetry remains.  The tight binding model is presented and their strengths are determined using the Slater-Koster theory for a given $(\phi,\theta)$.
In Sec.~3, we develop a microscopic Hamiltonian for the large Hubbard interaction, $U$, limit, and investigate Cooper pair instabilities within a mean field theory.
The transition between the odd-parity superconducting phase and the even-parity phase is presented while tuning the tilting and rotation angles, as well as the chemical potential in Sec.~4.
We summarize our findings and discuss experimental proposals to test our theory in the last section.

\section{Crystal Structure and Electronic Band Dispersion}

In a single layer of Sr$_2$IrO$_4$, each iridium atom is surrounded by six oxygen atoms, forming an octahedral cage.  Each cage creates a crystal field, which splits the $5d$ levels of the Ir atoms into triply degenerate $t_{2g}$ and doubly degenerate $e_g$ levels.  The relatively strong SOC of Ir then splits the $t_{2g}$ levels into $J_{\text{eff}}=\frac{1}{2}$ and $J_{\text{eff}}=\frac{3}{2}$, which leaves a half-filled $J_{\text{eff}}=\frac{1}{2}$ band when the 5 valence electrons of Ir$^{4+}$ are considered.  The SOC mixes the $d$-orbitals of the iridium atoms, and the two $J_{\text{eff}}=\frac{1}{2}$ states are defined as $|J_z = \pm \frac{1}{2} \rangle = \frac{1}{\sqrt{3}}(|d_{xy,\pm s}\rangle \pm |d_{yz,\mp s}\rangle +i|d_{xz,\mp s}\rangle)$, where $\pm s$ represents the $\pm \frac{1}{2}$ spin of the electron.

\begin{figure}
\includegraphics[width=\columnwidth]{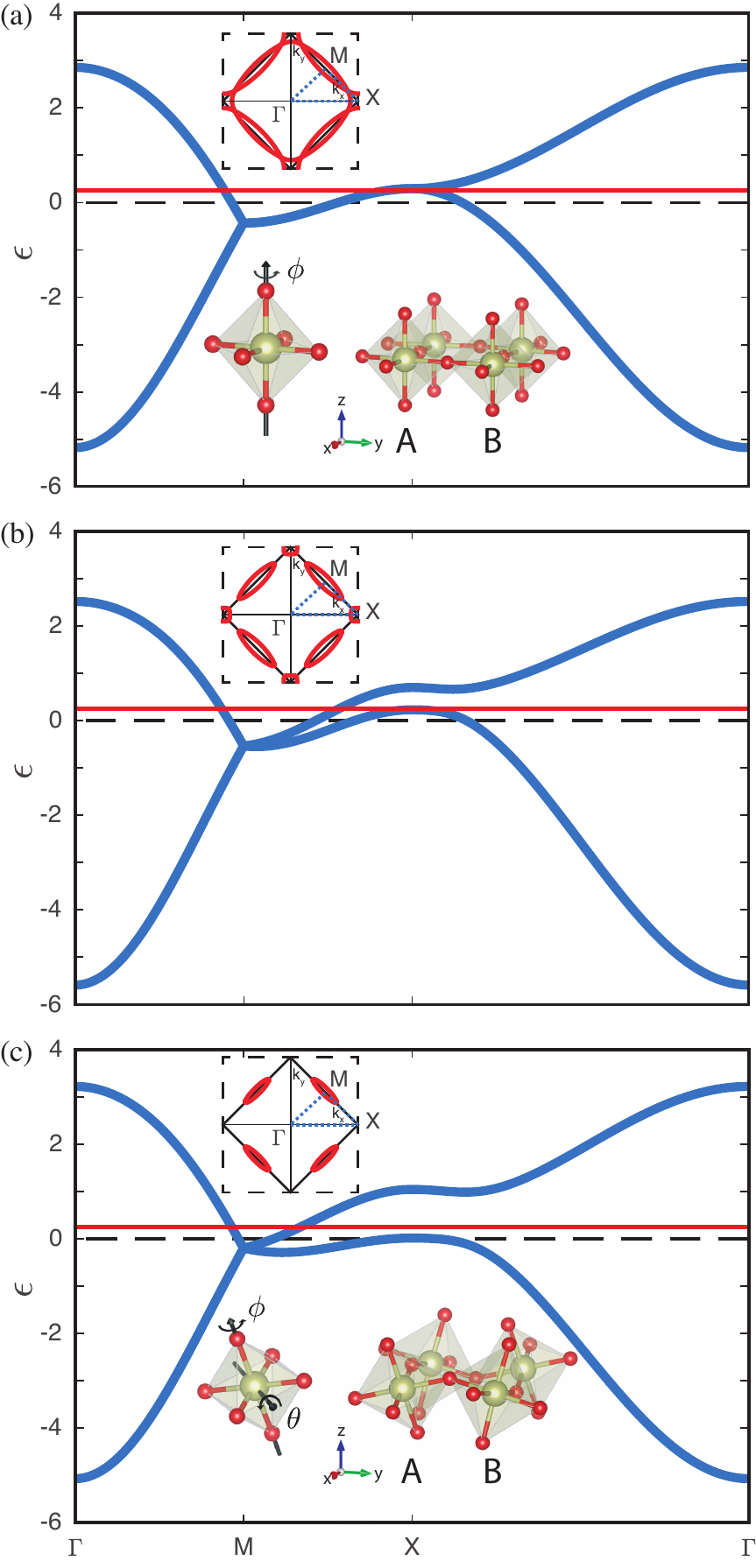}
\caption{Two-dimensional (2D) Band dispersion of $J_\text{eff}=\frac{1}{2}$ states, with (a) finite rotation ($\phi=11\degree$) only (i.e., no tilting), (b) small finite tilting ($\theta=5\degree$) and rotation ($\phi=5\degree$), and (c) finite tilting ($\theta=12\degree$) and rotation ($\phi=11\degree$).  A small chemical potential is shown as the solid, red line at $\mu \approx 0.2$ to represent electron doping.  Upper insets: First Brillouin zone (BZ) of 2D lattice with the $A$ and $B$ sublattices, and a dashed square original BZ without rotations.  The location of the Fermi surface is shown in red, corresponding to the small chemical potential shown in the band dispersion.  The path of the band dispersion plot is shown as the dotted blue line between points $\Gamma = (0,0)$, X $= (\pi, 0)$, and M $= (\pi/2, \pi/2)$.  Lower insets of (a) and (c):  (left) A single IrO$_6$ octahedron with a rotation of $\phi$ along the local $c$-axis, and a tilting of $\theta$ along the (110) direction.  (right) Four Ir sites, showing the two distinct sites, where NN sites having opposite rotations, i.e., $(\phi,\theta)$ and $(-\phi,-\theta)$.}\label{BS}
\end{figure}

The oxygen octahedra can be rotated about the local $z$-axis by an angle $\phi$, as shown in the lower inset of Fig.~\ref{BS}(a).  The rotation will alternate between nearest-neighbour (NN) sites, such that an iridium site with an octahedron rotated by an angle $\phi$ will have neighbours with a rotation angle $-\phi$.  In the presence of this rotation, the hopping integrals between the $d_{yz}$ and $d_{xz}$ orbitals will becomes finite, which will introduce a hopping term proportional to $\sigma_z$, where $\vec{\sigma}$ consists of the Pauli matrices, representing the $J_{\text{eff}}=\frac{1}{2}$ states.

The oxygen octahedra can also be tilted about the local $b$-axis by an angle $\theta$, as shown in the lower inset of Fig.~\ref{BS}(c), where the $a$-axis corresponds to the $(\bar{1}10)$ direction, the $b$-axis corresponds to the $(110)$ direction, and the $c$-axis corresponds to the $(001)$ direction.  This tilting also alternates between NN sites, such that an octahedron which is rotated and tilted by angles $(\phi,\theta)$, respectively, will have NN octahedra which are rotated and tilted by angles $(-\phi,-\theta)$.  Tilting is absent in bulk Sr$_2$IrO$_4$, but can be generated in an iridium oxide layer grown on a band insulator substrate, AMO$_3$ with P$_\text{bnm}$ crystal structure.  In the presence of alternating rotation and tilting of the octahedra between NN sites, the inter-orbital hopping integrals become finite between the $d_{yz}/d_{xz}$, and $d_{xy}$ orbitals.  In addition to the term proportional to $\sigma_z$ from hopping between the $d_{yz}$ and $d_{xz}$ orbitals, since the hopping of an electron of spin $s$ between the $d_{xy}$ and the $d_{yz}/d_{xz}$ orbitals involves hopping between the $|J_z = +\frac{1}{2} \rangle$ and $|J_z = -\frac{1}{2} \rangle$ states, a spin-flip term proportional to $\sigma_x$ or $\sigma_y$ will be present.

The unit cell consists of two square sublattices, labeled $A$ and $B$, each made up of iridium atoms inside of oxygen octahedra with the same tilting and rotation angles, such that all of one sublattice is characterized by rotation and tilting, $(\phi,\theta)$, while the other is completely characterized by $(-\phi,-\theta)$.  The tight-binding Hamiltonian can be expressed as
\begin{equation}\label{TBH}
\begin{split}
H_t= &\sum_{\langle i,j \rangle} \sum_{\sigma} t_0  c_{i A \sigma}^\dagger c_{j B \sigma} + i c_{i A \alpha}^\dagger (\vec{v} \cdot \vec{\sigma})_{\alpha \beta} c_{j B \beta}+ \text{h.c.} \\ &+ \sum_{\langle\langle i,j \rangle\rangle}\sum_{\sigma} t' (c_{iA\sigma}^\dagger c_{jA\sigma} +c_{iB\sigma}^\dagger c_{jB\sigma}) + \text{h.c.},
\end{split}
\end{equation}

where $\vec{v}=(\frac{1}{2} t_x,\frac{1}{2} t_y,t_z)$ along the $x$-axis, and $\vec{v}=(\frac{1}{2} t_y,\frac{1}{2} t_x,t_z)$ along the $y$-axis.  The hopping amplitudes are defined such that $t_0$ represents NN intra-orbital hopping, $t_z$, NN hopping between $d_{yz}$ and $d_{xz}$ orbitals, $t_x$ ($t_y$), NN hopping between $d_{xy}$ and $d_{xz}$ ($d_{yz}$) orbitals along the $x$-direction, and $t'$, next-nearest neighbour (NNN) intra-orbital hopping.  The hopping parameters can be determined for a given $(\phi,\theta)$ via the Slater-Koster theory \cite{Slater1954pr}.  In the presence of rotation only, $t_x=t_y=0$, while finite tilting ($\theta$) and rotation ($\phi$) will give all terms finite.  For example, taking the overlap of the $d$-orbitals as $(t_{dd\sigma},t_{dd\pi},t_{dd\delta})=(\frac{3}{2},-1,\frac{1}{4})$, and a suppression factor of $\alpha=0.4$ for NNN hopping, determined previously by matching a LDA band structure \cite{Carter2012prb}, where $(t_{dd\sigma}',t_{dd\pi}',t_{dd\delta}')=\alpha*(t_{dd\sigma},t_{dd\pi},t_{dd\delta})$, the parameters are listed for various rotation and tilting angles in Table \ref{SK}.  For the remainder of this paper, we will work in units where the largest hopping term, $|t_0|$, is set to 1.

\begin{table}
\centering
\begin{tabularx}{1\columnwidth}{|>{\centering}X |>{\centering}X |>{\centering}X |>{\centering}X |>{\centering}X |>{\centering}X|}
	\hline
	$(\phi,\theta)$ & $t_0$ & $t_x$ & $t_y$ & $t_z$ & $t'$ \tabularnewline \hline
	$(5\degree, 5\degree)$ & $-0.6$ & $0.15$ & $0.05$ & $-0.07$ & $-0.15$ \tabularnewline \hline
	$(11\degree, 12\degree)$ & $-0.6$ & $0.3$ & $0.1$ & $-0.15$ & $-0.1$ \tabularnewline \hline
	$(15\degree, 15\degree)$ & $-0.55$ & $0.3$ & $0.13$ & $-0.2$ & $-0.1$ \tabularnewline \hline
	$(20\degree, 20\degree)$ & $-0.5$ & $0.35$ & $0.17$ & $-0.25$ & $-0.1$ \tabularnewline \hline
	$(25\degree, 25\degree)$ & $-0.475$ & $0.35$ & $0.2$ & $-0.3$ & $-0.11$ \tabularnewline \hline
\end{tabularx}
\caption{Slater-Koster values of hopping parameters, $t_0,t_x,t_y,t_z$, and $t'$, for different rotation and tilting angles, $\phi$ and $\theta$, all defined in the main text.  Note the use of $(\phi,\theta)=(11\degree, 12\degree)$ is chosen to reflect the rotation and tilting found in bulk SrIrO$_3$ \cite{Zhao2008jap}.
}\label{SK}
\end{table}

The band structure is shown in the case of a finite rotation, but no tilting in Fig.~\ref{BS}(a).  This is comparable to the band structure used to predict $d$-wave superconductivity in Sr$_2$IrO$_4$ \cite{Bertinshaw2019}, based on its similarity to the cuprate superconductors other than the change in sign of the NNN hopping parameter \cite{Wang2011prl}.  
The Fermi surface in the presence of small electron doping consists of hole and electron pockets near the X$=(\frac{\pi}{2},\frac{\pi}{2})$ and M$= (\pi,0)$ points respectively, as shown in the upper inset of Fig.~\ref{BS}(a). 
By introducing a finite tilting angle, as shown in Figs.~\ref{BS}(b-c), the degeneracy between the X- and M-points is split by breaking the mirror symmetry in the $xy$-plane, leaving only Dirac points at the $(\pm \frac{\pi}{2},\pm \frac{\pi}{2})$ points, protected by a b-glide symmetry.  
 This b-glide symmetry consists of a reflection in the $bc$-plane at $a=1/4$, then translation along the $b$ axis. The b-glide operator interchanges $d_{xz}$ and $d_{yz}$ orbitals, in addition to exchanging the $A$ and $B$ sublattices.  Introducing the additional Pauli matrices, $\vec{\tau}$, to represent the sublattice, the b-glide symmetry operation can be expressed as,
$\hat{\Pi}_b=\frac{i}{\sqrt{2}}(\sigma_x - \sigma_y)\tau_x \hat{k}_{b}$,
where $\hat{k}_b$ is an operator acting on the crystal momentum such that $(k_x,k_y) \rightarrow (k_y,k_x)$ \cite{Carter2012prb}.  If this b-glide symmetry is broken, and the Dirac nodes are gapped, it was shown that this becomes a 2D topological insulator \cite{Chen2014prb}.  Leaving the b-glide symmetry intact, in the presence of small tilting and rotation, the Fermi surface in the presence of a small electron doping again consists of hole and electron pockets near the X- and M-point respectively, however, the pockets are not connected in this case, as shown in the upper inset of Fig.~\ref{BS}(b).  When the tilting and rotation angles are increased, only the electron pockets remain, as shown in the upper inset of Fig.~\ref{BS}(c).

Since formation of the superconducting gap involves the FS instability, the shape of the FS plays an crucial role in determining the pairing symmetry.  
In the case where the FS is limited to areas near the M-points, $d$-wave pairing of the form $\cos k_x - \cos k_y$ is likely unfavourable because the pairing gap is minimum around the M-points.  Alternatively, in the case of an odd parity pairing such as chiral, $p$-wave, of the form $\sin k_x \pm i\sin k_y$, the pairing gap is largest around the M-points, providing a possible reason 
why $p$-wave pairing may occur instead of $d$-wave.  In the next chapter, we will examine all possible pairing symmetries that arise in the large Hubbard $U$ interaction limit in the presence of finite $\theta$ and $\phi$.

\section{Superconducting instabilities in the large $U$ limit}
Starting with the Hubbard model,
\begin{equation}
H=H_t+U\sum_i n_{i\uparrow}n_{i\downarrow},
\end{equation}
and taking the large $U$ limit, we have the following extended spin model away from half filling.
\begin{multline}\label{largeUH}
H_\text{eff}=H_t+\sum_{i} \sum_{\delta\in[\pm x,\pm y]} [J \vec{S}_{i} \cdot \vec{S}_{i+\delta} + \vec{D}_{i,\delta} \cdot (\vec{S}_{i} \times \vec{S}_{i+\delta}) \\ - \tilde{V} n_{i}n_{i+\delta} + \Gamma^{\alpha \beta}_\delta S_{i}^\alpha S_{i+\delta}^\beta ] +\sum_{\langle\langle i,j \rangle\rangle} J'(\vec{S}_{i} \cdot \vec{S}_{j} - \frac{1}{4}n_{i}n_{j}).
\end{multline}
The interaction terms here are defined as the following \cite{Dzyaloshinsky1958,Moriya1960},
\begin{eqnarray}\label{ints}
J&=&\frac{4}{U}(t_0^2-|\vec{v}|^2), \quad
\vec{D}_{i,\delta}=\frac{8}{U} \epsilon_{i} t_0 \vec{v}_{\delta} \nonumber \\
\tilde{V}&=&\frac{1}{U}(t_0^2 + |\vec{v}|^2), \quad
\Gamma_\delta^{\alpha \beta} = \frac{8}{U}v_{\delta}^\alpha v_{\delta}^\beta \nonumber \\
J'&=&\frac{4 }{U}t'^2,
\end{eqnarray}  
where $\epsilon_i$ represents a change of sign between adjacent bonds in the DM term, $\vec{D}_{i,\delta}$, and the distinction between the $x$ and $y$ directions of $\vec{D}_{i,\delta}$ and $\Gamma_\delta^{\alpha \beta}$ reflects the difference of $\vec{v}$ along either the $x$ or $y$ direction.
The tight binding portion of the Hamiltonian, $H_t$, is supposed to project out hoppings that change the number of doubly occupied sites, but we find that that such projection does not affect the mean field results presented in the following section.

\begin{figure}
\includegraphics[width=\columnwidth]{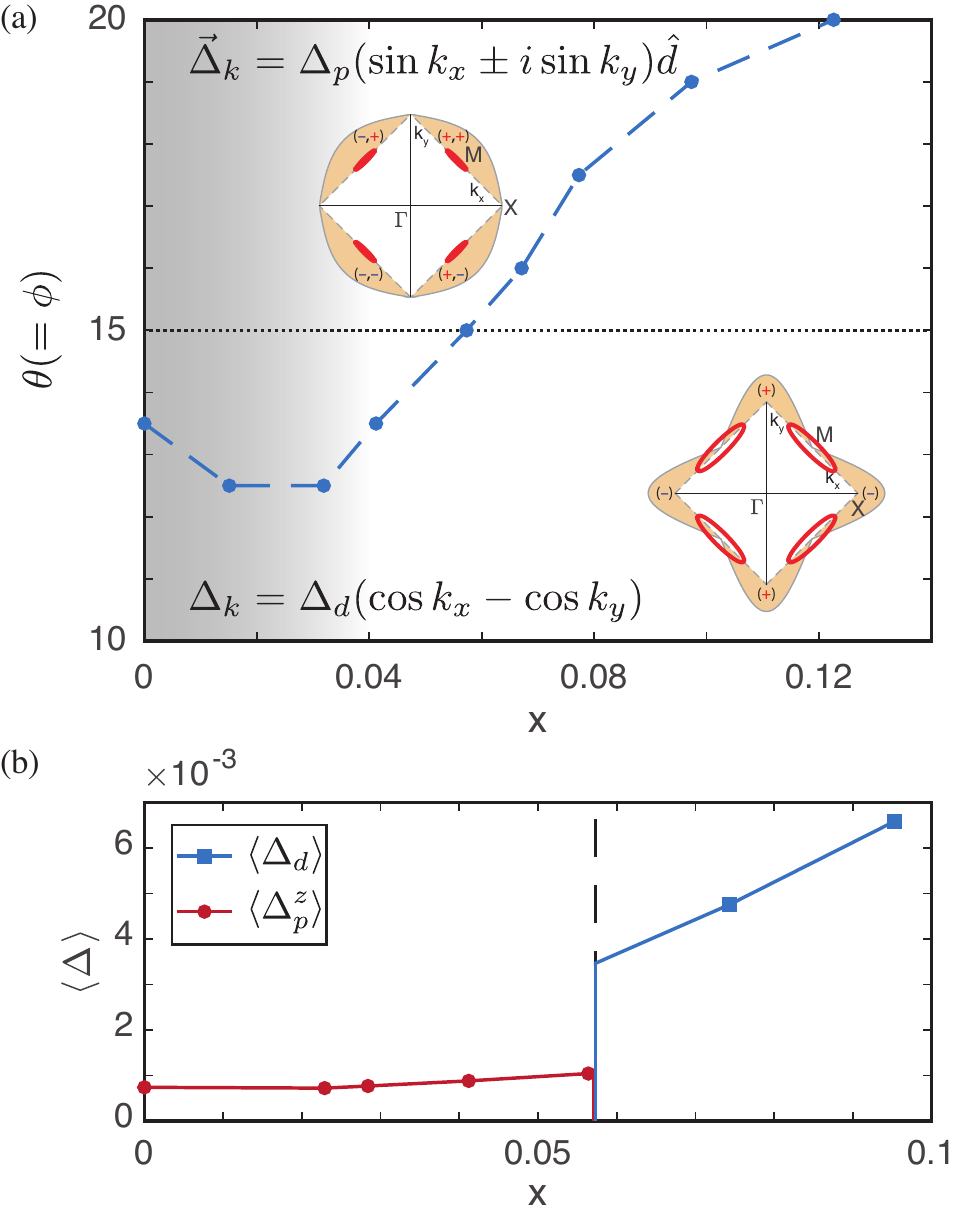}
\caption{(a) Phase diagram by varying the tilting angle $\theta$ and doping $x$.
 For small tilting angles, 
 the $d$-wave pairing is the minimum solution of free energy. However, $p$-wave becomes the minimum solution for larger angles, 
as indicated by the blue dashed phase transition line.  Both the $p$-wave and $d$-wave pairings are inset in their respective regions, plotted along the first Brillouin zone, 
 with a sample Fermi surface shown in red.  This shows that the $p$-wave pairing is largest for the FS near the M-point, while the $d$-wave paring is favoured as the FS elongates
 towards the X- and Y-points by increasing the electron doping, $x$.  A black dotted horizontal line is the path chosen for (b), to show the nature of the phase transition.
 The gray shaded area is where the magnetic ordering dominates and our pairing analysis may not be applicable. (b) Amplitudes of the order parameters for both the even- and odd-parity solutions for a fixed $(\theta,\phi)=(15\degree,15\degree)$, and varying doping,
$x$.  There are two sizeable order parameters (belonging to separate solutions), $\langle \Delta_p^z \rangle$  (red line with circles) and $\langle \Delta_d \rangle$ (blue line with squares), and all other order parameters are negligible.  The jump of the order parameter shows that the transition is first order (shown with the black, vertical line)}\label{PD} 
\end{figure}

Treating the effective Hamiltonian in mean field theory, we obtain the following mean field  Hamiltonian, 
\begin{multline}\label{HMF}
H_{\text{eff}}^{\text{MF}} =H_t+\sum_\delta \bigg\{ \frac{N}{V_S}\langle \Delta_{s_\delta}^{\dagger}\rangle\Delta_{s_\delta} + \sum_\alpha \bigg[ \frac{N}{V_{T_\delta}^\alpha}\langle\Delta_{p_\delta}^{\alpha\dagger}\rangle \Delta_{p_\delta}^\alpha \\- \sum_{\beta \ne \alpha} \frac{4N}{\Gamma_\delta^{\alpha \beta}}\langle\Delta_{p_\delta}^{\alpha\dagger}\rangle\Delta_{p_\delta}^\beta  \bigg] \bigg\}+ \text{h.c.} + \text{const.,}
\end{multline}
where $\alpha \in [x, y, z]$ labels the three spin-triplet components of the triplet order parameters.  The order parameter operators are defined as,
\begin{eqnarray}\label{delta}
\Delta_{s_x} &=&\frac{V_S}{N}\sum_{\bf k} \psi_{{\bf k} \alpha} (i\sigma_y)_{\alpha \beta} (\tau_x) \psi_{-{\bf k} \beta}(\cos{k_x}) \nonumber\\
\Delta_{s_y}&=& \frac{V_S}{N}\sum_{\bf k} \psi_{{\bf k} \alpha} (i\sigma_y)_{\alpha \beta} (\tau_x) \psi_{-{\bf k} \beta}(\cos{k_y}) \nonumber\\
{\Delta}^\alpha_{p_x} &=& \frac{V_{T_x}^\alpha}{N}\sum_{\bf k} \psi_{{\bf k} \alpha} ({\hat d}_\alpha \sigma_\alpha i\sigma_y)_{\alpha \beta}(\tau_x)\psi_{-{\bf k} \beta}(\sin{k_{x}}) \nonumber\\
{ \Delta}^\alpha_{p_y} &=&\frac{V_{T_y}^\alpha}{N}\sum_{\bf k} \psi_{{\bf k} \alpha} ({\hat d}_\alpha \sigma_\alpha i\sigma_y)_{\alpha \beta}(\tau_x)\psi_{-{\bf k} \beta}(\sin{k_y}),
\end{eqnarray}
representing the spin singlet ($\Delta_s$) and triplet ($\Delta_p^\alpha$) with $\alpha = x, y, z$, and $\psi_{{\bf k} \sigma}=(c_{A {\bf k} \sigma},c_{B{\bf k} \sigma})^T$.  Note that since inversion symmetry is not broken, the DM term does not couple the singlet and triplet order parameters as is the case when the SOC breaks inversion symmetry \cite{Gorkov2001prl}.
Given that $x$- and $y$-axis rotational symmetry is broken, $x$- and $y$-bond pairings 
 can be different.
The extended $s$-wave and $d$-wave pairings are represented by  $\Delta_s (\cos k_x + \cos k_y)$
and  $\Delta_d (\cos k_x - \cos k_y)$, respectively and $\Delta_{s/d} \equiv \langle \Delta_{s_x} \rangle = \langle \Delta_{s_y} \rangle$.
 For the $p$-wave pairing ${\vec \Delta}_p$, the $d$-vector direction is pinned by
the interaction $V_T^\alpha$. In other words, the $\alpha$-direction of the $d$-vector originates from the interaction $V_T^\alpha$ shown below.
The effective interaction terms are given by,
\begin{eqnarray}\label{V}
V_{S} &=& -\frac{3J}{8}-\frac{\tilde{V}}{2}-\frac{1}{8}(\Gamma^{xx}+\Gamma^{yy}+\Gamma^{zz}) \nonumber\\
V_{T_\delta}^\alpha &=& -\frac{\Gamma_\delta^{\alpha \alpha}}{4} + \bigg[ \frac{J}{8}-\frac{\tilde{V}}{2} +\frac{1}{8}(\Gamma^{xx}+\Gamma^{yy}+\Gamma^{zz})\bigg].
\end{eqnarray}
Here, $\Gamma^{\alpha \alpha}$, induced from the finite tilting, generates an attractive interaction, $V_T^\alpha$, for the spin triplet pairing
as well as the spin singlet pairing, while the conventional Heisenberg leads to an attractive interaction for only the spin singlet. 
Below, we present the self-consistent mean-field results as we vary the rotation and tiling angles of the octahedra, as well as the electron doping level.

\section{Transition between even and odd parity pairings}

The gap equations can be solved in the presence of the interaction terms of Eq.~(\ref{largeUH}) at zero temperature.  This was done in the presence of only the intra-orbital hopping term, and is discussed in Ref.~\onlinecite{Coleman2015}.
Considering the case of only finite rotations as shown in Fig.~\ref{BS}(a), including only the Heisenberg and density-density terms leads to a $d$-wave singlet.

When a finite local tilting is introduced, as shown in Fig.~\ref{BS}(c), the $x$- and $y$-components of the $\Gamma$ term are finite, and an odd-parity dominant solution can occur due to the attractive triplet interactions as well as the location of the Fermi surface.  The order parameters are solved self-consistently for various values of electron doping, representing the shift of the Fermi surface away from the $(\pm \frac{\pi}{2},\pm \frac{\pi}{2})$ points, as well as various tilting and rotation angles, which change the interaction strength of the triplet terms relative to the singlet.  In addition to the $d$-wave solution discussed above, an odd-parity, triplet solution can occur in the form of ${\vec \Delta} ({\bf k})  =\Delta_p {\hat d} \sin k_x$,  $\Delta_p {\hat d} \sin k_y$, or a linear combination of the two
such as a chiral solution of $\Delta_p {\hat d} (\sin k_x \pm i \sin k_y)$, where $\Delta_p = \langle \Delta_{p_x} \rangle  = \langle {\Delta}_{p_y} \rangle$
and direction of $d$-vector ${\hat d}$ can be pinned by the dominant $\Gamma^{\alpha\alpha}$ interaction in Eq.~(\ref{V}).

By minimizing the free energies with respect to 16 mean field order parameters, including both real and imaginary components,  
the state corresponding to the free energy minimum can be found.  The phase diagram generated by varying
the tilting angle, $\theta $, (we set the rotation angle, $\phi$, to be the same, but different values of $\phi$ do not change the qualitative result) 
and electron doping, $x$, is shown in Fig.~\ref{PD}(a).  To obtain the phase diagram, we used a set of tight binding parameters for a given set of angles, 
and a few examples are listed in Table \ref{SK}.  The effective interaction values are estimated from the tight-binding parameters, as described in Eq.~(\ref{ints}), with $U=2.3t_0$.  Various critical values of the angles are found for different doping concentrations.
%

Below the critical angle for a given doping, the $d$-wave dominant solution occurs.  The $d$-wave solution occurs for small doping mainly
 due to hole pockets in the Fermi surface near the X- and Y-points. However, for small $x$, the magnetic ordering is dominant, as indicated by the shaded region in Fig.~\ref{PD}(a),
 and thus the $d$-wave superconducting state would appear only at higher doping and small tilting, $\theta$.  The $d$-wave pairing at higher doping
 is achieved via the energy gain by gapping out the FS near the X- and Y-points due to the elongated FS shown by the red line in the lower inset of Fig.~\ref{PD}(a). 

Above the critical angle, the odd-parity dominant solutions indeed occurs, as the small pockets are near the M-point, 
and for higher doping the odd-parity state requires a higher tilting angle.
The blue dashed line represents the first order phase transition between the odd- and even-pairing states.
The precise transition line may change
 depending on the details of tight binding parameters, while the qualitative behaviour does not depend on the details.
%
 The direction of the $d$-vector in the $p$-wave pairing is determined by the interaction term, $\Gamma^{\alpha\alpha}$.  
 For all cases we considered, the $d$-vector is pinned along the ${\hat z}$-direction indicating
 that $\Gamma^{xx}, \Gamma^ {yy} < \Gamma^{zz}$.
 There are 4 degenerate solutions found in the odd-parity state; one is $\vec{d} ({\bf k}) \propto (\sin k_x) \hat{z}$, another is $\vec{d} ({\bf k}) \propto (\sin k_y) \hat{z}$, and
 the other two are
$\vec{d} ({\bf k}) \propto (\sin{k_x} \pm i \sin{k_y})\hat{z}$,
which is a particular linear combination of the other  two. 
Note that this requires imaginary $\pm i  \sin k_y$, and thus the state spontaneously breaks time reversal symmetry by choosing either $+i$ or $-i$.

In Fig.~\ref{PD}(a),  we present only which state corresponds to the free energy minimum.  To show how each order parameter changes as a function of the doping $x$, we choose
a specific angle, represented by the dashed horizontal line at $(\phi,\theta)=(15\degree,15\degree)$ in Fig.~\ref{PD}(a), 
and show the magnitudes of the different components  in Fig.~\ref{PD}(b).  
There are two sizeable order parameters, $\langle \Delta_p^z \rangle$ indicated by the red line with squares, and $\langle \Delta_d \rangle$ indicated by the blue line with circles.
All other order parameters including the extended s-wave and x- and y-component of triplet pairings are negligible.  While the magnitude of the $d$-wave order parameter is almost always larger, it does not represent the free energy minimum in the phase diagram because it is suppressed by the form factor, which is small near the Fermi surface for small doping values.
The order parameter jumps from $p$-wave to $d$-wave at the transition, indicating a first order transition. The precise location of transition point depends on the details of tight binding parameters, while the qualitative main finding of the transition from the $d$-wave to the $p$-wave pairing does not. 

\section{Summary and Discussion}

The bulk Sr$_2$IrO$_4$ has offered a playground to study combined effects of SOC and electronic correlations.
For example, the Mott insulator is originated from the Hubbard $U$ interaction on the effectively narrow $J_{\text{eff}}=\frac{1}{2}$ SOC band.
Based on several similarities to high-T$_c$ cuprates, including a single band basis, 
strong electron-electron interaction, and crystal structure, the $d$-wave superconducting pairing
was proposed for electron doped Sr$_2$IrO$_4$ \cite{Wang2011prl,Watanabe2013prl,Yang2014prb,Meng2014prl}.
Despite a lack of experimental evidence of a bulk superconducting state,  the $d$-wave-like nodal gap was observed by ARPES on surface doped Sr$_2$IrO$_4$ \cite{Kim2016np}.

On the other hand, ARPES data collected in electron doped samples using La atoms on a superlattice made of [SrIrO$_3$,SrTiO$_3$]  has shown evidence of the Dirac point, as well as Fermi pockets forming around the $(\pm \frac{\pi}{2},\pm \frac{\pi}{2})$ points \cite{Brouet2015prb,DelaTorre2015prl}.
This indicates that such a superlattice with La doping has a qualitatively different FS from the surface doped Sr$_2$IrO$_4$.  
It is plausible that the La doping in this superlattice introduces further octahedra distortion, and 
the $p$-wave pairing symmetries may occur via the octahedra distortions generated by superlattice and/or under a strain. Thus, tuning of the octahedra distortions is a tool for engineering a $p$-wave superconductor.
Since the $p$-wave spin triplet is sensitive to disorder, it requires careful preparation and further analysis to determine the existence of
superconductivity in these superlattices. 

In summary, we studied effects of octahedra distortions on 
possible superconducting pairing symmetries in strongly spin-orbit coupled 2D iridium oxides with electron doping.
We found that with a finite staggered tilting of octahedra about the $[110]$-direction in addition to the staggered rotation about the $c$-axis, 
the FS changes to a Dirac point around the M-point. 
Such tilting also generates effective spin interactions in addition to the conventional Heisenberg interaction in large Hubbard $U$ limit.
These two changes -- FS topology and additional spin interactions -- originating from the octahedra tilting, 
lead to the phase transition from the $d$-wave spin singlet to $p$-wave spin triplet, as the tilting angle increases.  While the above analysis was made for iridium oxides, the main idea of the combined effects of SOC and lattice structures can be applied to other systems with strong SOC.  Effects of disorder and other competing states would be important subjects for future studies.


\bibliography{SC-SOC}

\end{document}